\documentclass[12pt]{article}
\usepackage{axodraw,bbold}

\catcode`@=12
\begin{document}
\thispagestyle{empty}
\begin{flushright} 
UCRHEP-T419\\ 
September 2006\
\end{flushright}
\vspace{0.5in}
\begin{center}
{\Large	\bf Emanations of Dark Matter: 
Muon Anomalous\\ Magnetic Moment, 
Radiative Neutrino Mass,\\
and Novel Leptogenesis at the TeV Scale\\}
\vspace{1.5in}
{\bf Thomas Hambye$^a$, Kristjan Kannike$^b$, Ernest Ma$^c$, and Martti 
Raidal$^b$\\}
\vspace{0.2in}
{$^a$ \sl Instituto F\'{i}sica Te\'{o}rica, Universidad Aut\'{o}noma de 
Madrid, Cantoblanco, Spain\\}
\vspace{0.1in}
{$^b$ \sl National Institute of Chemical Physics and Biophysics, Ravala 10, 
Tallinn, Estonia\\}
\vspace{0.1in}
{$^c$ \sl Physics Department, University of California, Riverside, 
California 92521, USA\\}
\vspace{1.0in}
\end{center}

\begin{abstract}\
The evidence for dark matter signals a new class of particles at the TeV 
scale, which may manifest themselves indirectly through loop effects. In a 
simple model we show that these loop effects may be responsible for the 
enhanced muon anomalous magnetic moment, for the  
neutrino mass, as well as for leptogenesis in a novel way. This 
scenario can be verified at LHC experiments.
\end{abstract}

\newpage
\baselineskip 24pt

There are at present two solid pieces of evidence for physics beyond the 
standard model (SM) of particle interactions. One is neutrino mass and 
the other is dark matter.  As pointed out recently \cite{m06-1,kms06,m06-2,
m06-3,nudm}, the two may be intimately related.  There is however another hint 
for physics beyond the SM, i.e. the muon anomalous magnetic moment 
\cite{g-2}.  It may also be related to neutrino mass, as pointed out 
some time ago \cite{mr01,r01}.  Here we show how all three
may be connected and allow in addition \underline{novel} realistic 
leptogenesis \cite{lepto} at the TeV scale.  We assume that the particles 
responsible for the \underline{strongly enhanced} muon anomalous magnetic 
moment are all members of a new class of particles which 
are {\it odd} under an exactly conserved discrete $Z_2$ symmetry, whereas all 
SM particles are even.  The lightest particle of this class is 
absolutely stable and, assuming that it is also neutral, it becomes a 
good candidate for the cold dark matter of the Universe \cite{dm_rev}.  
In particular we propose a specific minimal model where neutrinos also 
obtain small radiative Majorana masses from exactly the same particles.

\begin{table}[htb]
\caption{Particle content of proposed model.}
\begin{center}
\begin{tabular}{|c|c|c|c|c|}
\hline 
particles & $SU(2) \times U(1)$ & $U(1)_L$ & $(-1)^L$ 
& $Z_2$ \\ 
\hline
$L_\alpha = (\nu_\alpha,l_\alpha)$ & $(2,-1/2)$ & $1$ & $-$ & + \\ 
$l^c_\alpha$ & $(1,1)$ & $-1$ & $-$ & + \\ 
$\Phi = (\phi^+,\phi^0)$ & $(2,1/2)$ & 0 & + & + \\ 
\hline
$N_i$ & $(1,0)$ & $1$ & $-$ & $-$ \\ 
$N^c_i$ & $(1,0)$ & $-1$ & $-$ & $-$ \\ 
$\eta = (\eta^+,\eta^0)$ & $(2,1/2)$ & 0 & + & $-$ \\ 
$\chi^-$ & $(1,-1)$ & 0 & + & $-$ \\ 
\hline
\end{tabular}
\end{center}
\end{table}

Consider the particles listed in Table 1.  There are of course the SM 
lepton doublets $L_\alpha$ and singlets $l^c_\alpha$, as well as the usual 
Higgs doublet $\Phi$ with $\langle \phi^0 \rangle = v$.  The neutral leptons 
$N_i,N^c_i$ are new, as well as the scalar doublet $(\eta^+,\eta^0)$ and 
singlet $\chi^-$.  They transform nontrivially under the global lepton
$U(1)_L$ and the $Z_2$ symmetries as indicated. Although Table 1 
contains several new particles, they are optimized to 
encompass the various otherwise disparate issues of dark matter, muon 
anomalous magnetic moment, neutrino mass, 
as well as leptogenesis at the TeV scale.
Assuming the conservation 
of $U(1)_L$ and $Z_2$, the relevant allowed terms in the Lagrangian
involving these particles are given by
\begin{eqnarray}
{\cal L} &=& f_\alpha (\nu_\alpha \phi^- + l_\alpha \bar \phi^0) l^c_\alpha 
+ h_{\alpha i} (\nu_\alpha \eta^0 - l_\alpha \eta^+) N^c_i + h'_{\alpha i} 
l^c_\alpha \chi^- N_i \nonumber \\ &+& M_i N_i N^c_i + \mu (\phi^+ \eta^0 - 
\phi^0 \eta^+) \chi^- + {1 \over 2} \lambda_5 (\eta^\dagger \Phi)^2 + H.c.,
\end{eqnarray}
where $\alpha$ refers to the $(l_\alpha,l^c_\alpha)$ diagonal basis, i.e. 
$e,\mu,\tau$ mass eigenstates, and $i$ refers to the $(N_i,N^c_i)$ diagonal 
basis.  Note that $N^c_i$ is the Dirac mass partner of $N_i$.  Note also that 
neutrinos are massless at tree level because $\langle \eta^0 \rangle = 0$.

Consequently, there is a direct magnetic-dipole transition from $\mu$ to 
$\mu^c$, given by the diagrams of FIG.~1, involving all the  
new particles of Table 1 in the loop.
  
\begin{figure}[htb]
\begin{center}\begin{picture}(500,100)(10,45)
\ArrowLine(70,50)(110,50)
\ArrowLine(150,50)(190,50)
\ArrowLine(150,50)(110,50)
\ArrowLine(230,50)(190,50)
\Text(90,35)[b]{$\mu$}
\Text(210,34)[b]{$\mu^c$}
\Text(130,32)[b]{$N_i^c$}
\Text(170,33)[b]{$N_i$}
\Text(155,97)[b]{$\chi^-$}
\Text(105,70)[b]{$\eta^+$}
\Text(198,72)[b]{$\chi^-$}
\Text(111,112)[b]{$\phi^0$}
\Text(193,116)[b]{$\gamma$}
\DashArrowLine(130,85)(115,111){3}
\Photon(170,85)(185,111){2}{3}
\DashArrowArc(150,50)(40,120,180){3}
\DashArrowArcn(150,50)(40,60,0){3}
\DashArrowArcn(150,50)(40,120,60){3}

\ArrowLine(270,50)(310,50)
\ArrowLine(350,50)(390,50)
\ArrowLine(350,50)(310,50)
\ArrowLine(430,50)(390,50)
\Text(290,35)[b]{$\mu$}
\Text(410,34)[b]{$\mu^c$}
\Text(330,32)[b]{$N_i^c$}
\Text(370,33)[b]{$N_i$}
\Text(355,97)[b]{$\eta^+$}
\Text(305,70)[b]{$\eta^+$}
\Text(400,72)[b]{$\chi^-$}
\Text(310,112)[b]{$\gamma$}
\Text(395,112)[b]{$\phi^0$}
\Photon(330,85)(315,111){2}{3}
\DashArrowLine(370,85)(385,111){3}
\DashArrowArc(350,50)(40,120,180){3}
\DashArrowArc(350,50)(40,60,120){3}
\DashArrowArcn(350,50)(40,60,0){3}

\end{picture}
\end{center}
\caption[]{Dominant contributions to muon anomalous magnetic moment.}
\end{figure}
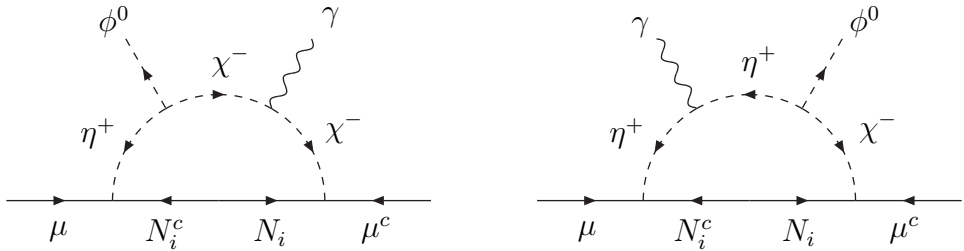

\noindent There are also the indirect transitions given by the diagrams of 
FIG.~2.  They are however expected to be small because they have an extra 
factor of $m_\mu$. Note that the existence of a muon anomalous magnetic 
moment is 
consistent with the conservation of $U(1)_L$ as well as the new $Z_2$.

\begin{figure}[htb]
\begin{center}\begin{picture}(500,100)(10,45)
\ArrowLine(70,50)(100,50)
\ArrowLine(130,50)(100,50)
\ArrowLine(130,50)(190,50)
\ArrowLine(220,50)(190,50)
\Text(85,35)[b]{$\mu$}
\Text(115,34)[b]{$\mu^c$}
\Text(205,34)[b]{$\mu^c$}
\Text(160,33)[b]{$N_i$}
\Text(130,70)[b]{$\chi^-$}
\Text(198,70)[b]{$\chi^-$}
\Text(102,95)[b]{$\phi^0$}
\Text(160,115)[b]{$\gamma$}
\DashArrowLine(100,50)(100,90){3}
\Photon(160,80)(160,110){2}{3}
\DashArrowArcn(160,50)(30,180,90){3}
\DashArrowArcn(160,50)(30,90,0){3}

\ArrowLine(280,50)(310,50)
\ArrowLine(370,50)(310,50)
\ArrowLine(370,50)(400,50)
\ArrowLine(430,50)(400,50)
\Text(295,35)[b]{$\mu$}
\Text(385,35)[b]{$\mu$}
\Text(415,34)[b]{$\mu^c$}
\Text(340,32)[b]{$N_i^c$}
\Text(310,70)[b]{$\eta^+$}
\Text(375,70)[b]{$\eta^+$}
\Text(340,115)[b]{$\gamma$}
\Text(402,95)[b]{$\phi^0$}
\Photon(340,80)(340,110){2}{3}
\DashArrowLine(400,50)(400,90){3}
\DashArrowArc(340,50)(30,90,180){3}
\DashArrowArc(340,50)(30,0,90){3}

\end{picture}
\end{center}
\caption[]{Subdominant contributions to muon anomalous magnetic moment.}
\end{figure}
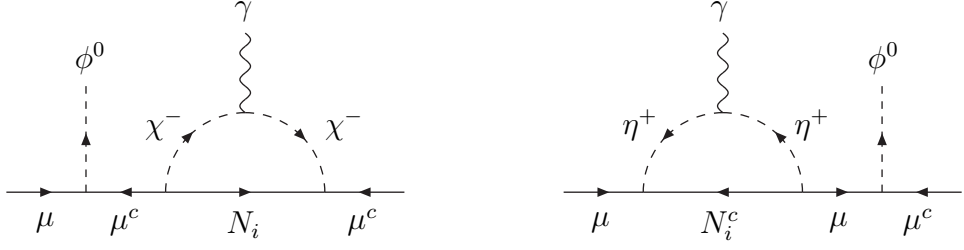

The diagrams of FIG.~1 can be computed exactly by using the mass 
eigenstates
\begin{eqnarray}
X^+ &=& \chi^+ \cos \theta  - \eta^+ \sin \theta, \\ 
Y^+ &=& \chi^+ \sin \theta + \eta^+ \cos \theta,
\end{eqnarray}
where
$\sin \theta \cos \theta (m_X^2 - m_Y^2) = \mu v. $
Their contribution to the muon anomalous magnetic moment 
$a_\mu = (1/2)(g-2)_\mu$ is then given by
\begin{equation}
\Delta a_\mu = {-\sin \theta \cos \theta \over 16 \pi^2} \sum_i 
h_{\mu i} h'_{\mu i} {m_\mu \over M_i} [F(x_i)-F(y_i)],
\label{gminustwo}
\end{equation}
where $x_i = m^2_X/M^2_i$, $y_i = m^2_Y/M^2_i$, and
\begin{equation}
F(x) = {1 \over (1-x)^3} [1 - x^2 + 2x \ln x].
\end{equation}
Let $y_i << x_i \simeq 1$, $M_i \sim 1$ TeV, and $(-h_{\mu i} h'_{\mu i} 
\sin \theta \cos \theta/24 \pi^2) \sim 10^{-5}$, we then obtain 
$\Delta a_\mu \sim 10^{-9}$, which compares favorably with the experimental 
value of \cite{g-2}
\begin{equation}
\Delta a_\mu = [(22.4 \pm 10)~{\rm to}~(26.1 \pm 9.4)] \times 10^{-10}.
\label{exp}
\end{equation}
Because of the chirality flip in the internal fermion line,  
$\Delta a_\mu$ in our model is 
strongly enhanced  by the factor ${\cal O}(v/m_\mu)$ compared to the 
usual result where the chirality flip occurs in the external muon line. 
In supersymmetry \cite{mw01},
\begin{eqnarray}
\Delta a_\mu^{\rm SUSY} &=& {\tan \beta \over 192 \pi^2} {m^2_\mu \over 
M^2_{\rm SUSY}} (5 g_2^2 + g_1^2) \nonumber \\ 
&=& 14 \tan \beta \left( {100~{\rm GeV} \over M_{\rm SUSY}} \right)^2 
10^{-10},
\label{susy}
\end{eqnarray}
thus requiring a large $\tan \beta$ or a small $M_{\rm SUSY}$ or both.  If 
$M_{\rm SUSY}$ turns out to be of order 1 TeV, it may be difficult to 
reconcile Eq.~(\ref{susy}) with Eq.~(\ref{exp}); whereas 
in our case, the experimental result, i.e. Eq.~(\ref{exp}),
can be obtained naturally with relatively small Yukawa 
couplings $h,\,h' \sim {\cal  O}(10^{-1}-10^{-2}). $

As for dark matter, the usual $R$ parity is identified as the new $Z_2$ of 
this model.  The lightest particle which is odd under $Z_2$ can be $Re \eta^0$ 
or $Im \eta^0$ with mass $m_0 \sim 60$ to 80 GeV and mass splitting of a few 
GeV as discussed in Ref.~\cite{bhr06}. This fixes the $\lambda_5$ coupling 
to be ${\cal O}(10^{-2})$.

Consider now the \underline{soft} explicit breaking of $U(1)_L$ down to its 
discrete subgroup $(-1)^L$ by the \underline{small}  Majorana mass terms
\begin{equation}
{1 \over 2} m_{ij} N^c_i N^c_j + {1 \over 2} m'_{ij} N_i N_j + H.c.
\label{softmasses}
\end{equation}
Neutrinos are still massless at tree level, but they may now acquire 
radiative masses in one loop as shown in FIG.~3.

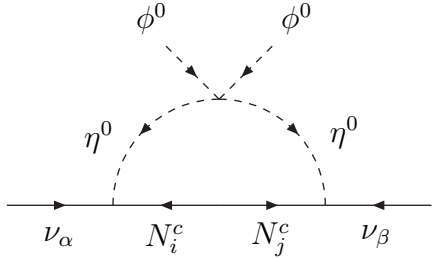
\begin{figure}[htb]
\begin{center}\begin{picture}(300,100)(10,45)
\ArrowLine(70,50)(110,50)
\ArrowLine(150,50)(190,50)
\ArrowLine(150,50)(110,50)
\ArrowLine(230,50)(190,50)
\Text(90,35)[b]{$\nu_\alpha$}
\Text(210,34)[b]{$\nu_\beta$}
\Text(130,32)[b]{$N_i^c$}
\Text(170,31)[b]{$N_j^c$}
\Text(105,70)[b]{$\eta^0$}
\Text(198,72)[b]{$\eta^0$}
\Text(125,115)[b]{$\phi^0$}
\Text(180,115)[b]{$\phi^0$}
\DashArrowLine(130,110)(150,90){3}
\DashArrowLine(170,110)(150,90){3}
\DashArrowArc(150,50)(40,90,180){3}
\DashArrowArcn(150,50)(40,90,0){3}

\end{picture}
\end{center}
\caption[]{Radiative Majorana neutrino mass.}
\end{figure}

\noindent As a result,
\begin{equation}
({\cal M}_\nu)_{\alpha \beta} = \sum_{i,j} h_{\alpha i} h_{\beta j} 
\tilde m_{ij},
\label{mnu}
\end{equation}
where $\tilde m_{ij} = |2 \lambda_5 v^2 m_{ij} I_{ij}|$, and
\begin{eqnarray}
I_{ij} &=& \int {d^4 k \over (2\pi)^4} {k^2 \over (k^2-m_0^2)^2} {1 \over (k^2 
- M_i^2)} {1 \over (k^2 - M_j^2)} \nonumber \\ 
&=& {1 \over 16 \pi^2 i (M_i^2-M_j^2)} \left[ {M_i^2 \over m_0^2-M_i^2} + 
{M_i^4 \ln (M_i^2/m_0^2) \over (m_0^2-M_i^2)^2} - (i \leftrightarrow j) 
\right].
\end{eqnarray}
\noindent Note that this expression for neutrino mass is not of 
the canonical seesaw form \cite{seesaw}. Parametrically it is
suppressed by two powers of the heavy scale rather than 
one, and numerically it involves
two extra suppression factors:
the $\lambda_5/16 \pi^2$ loop 
factor and the small $m/M$ factor.
The latter is due to the fact that the Majorana neutrino masses of this model 
are associated with the low breaking scale of global lepton number, i.e. 
$m_{ij}$ which are small because they are explicit symmetry breaking terms. The 
situation is similar to that of the supersymmetric models of 
Ref.~\cite{AHSMW}, but with renormalizable interactions.
These successive suppressions allow naturally 
small neutrino masses to be generated at the TeV scale
with ${\cal O}(10^{-2}-1)$ Yukawa couplings, leading to 
large production cross sections at colliders and to a 
large $\Delta a_\mu$ contribution, i.e. Eq.~(\ref{gminustwo}).  
Numerically, let 
$M_{i,j} \sim 1$ TeV, $m_{ij} \sim 0.1$ GeV, $h_{\alpha i} \sim 
{\cal O}(10^{-2})$, 
$\lambda_5 \sim {\cal O}(10^{-2})$, and $m_0 \sim v \sim 10^2$ GeV, then the 
entries of ${\cal M}_\nu$ are typically of order 0.1 eV. 
As a specific example, consider the possibility that $(N_1,N^c_1)$ have 
zero Yukawa couplings and let 
\begin{equation}
h_{\alpha i} \simeq h \pmatrix{0 & 1 & 0 \cr 0 & 0 & 1/\sqrt 2 \cr 
0 & 0 & 1/\sqrt 2},
\label{h}
\end{equation}
then in the basis $[\nu_e, (\nu_\mu+\nu_\tau)/\sqrt 2, (-\nu_\mu+\nu_\tau)/
\sqrt 2]$, Eq.~(\ref{mnu}) becomes
\begin{equation}
{\cal M}_\nu \simeq h^2 \pmatrix{\tilde m_{22} & \tilde m_{23} & 0 \cr 
\tilde m_{23} & \tilde m_{33} & 0 \cr 0 & 0 & 0},
\end{equation}
which is a realistic representation of present neutrino-oscillation data, 
with $\theta_{23} = \pi/4$, $\theta_{13} = 0$, and an inverted ordering of 
neutrino masses, having $m_3 = 0$.  Using Eq.~(\ref{h}), we see also that 
$\mu \to e \gamma$ is easily suppressed if we choose $h'_{\mu 2} \simeq 0$. 

To have successful leptogenesis at the TeV scale, we assume $(N_1,N^c_1)$ 
to be the lightest among the 3 pairs of singlet neutral fermions, and 
$h_{\alpha 1}, h'_{\alpha 1}$ to be very small, i.e. of 
order $10^{-7}$ instead of zero.  
This allows the decay of $(N_1,N^c_1)$ to satisfy the out-of-equilibrium 
condition, but will not affect Eq.~(\ref{h}), as far as ${\cal M}_\nu$ and 
$\Delta a_\mu$ are concerned. Because $\Delta m_{atm}^2$ 
and  $\Delta m_{sol}^2$ 
are induced by the two heavier generations $(N_{2,3},N^c_{2,3})$, 
this scenario predicts the lightest neutrino to be almost massless.

Consider first the $2 \times 2$ mass matrix spanning $N_1$ and $N_1^c$ and 
rotate it by $\pi/4$, i.e.
\begin{equation}
{1 \over \sqrt 2} \pmatrix{1 & -1 \cr 1 & 1} \pmatrix{m'_{11} & M_1 \cr M_1 
& m_{11}} {1 \over \sqrt 2} \pmatrix{1 & 1 \cr -1 & 1} = \pmatrix{-M_1 + 
A & B \cr B & M_1 + A},
\label{mass}
\end{equation}
where $A = (m'_{11}+m_{11})/2$, $B = (m'_{11}-m_{11})/2$.  Choose phases so 
that $M>0$, $A>0$ are real and $B = |B|e^{i \alpha}$.  Let the above mass 
matrix be diagonalized by
\begin{equation}
\pmatrix{ c e^{i \beta} & -s \cr s & c e^{-i \beta}}
\end{equation}
on the left and its transpose on the right, where $c = \cos \theta$ and 
$s = \sin \theta$.  Then
\begin{equation}
\sin \beta = {-M_1 \tan \alpha \over \sqrt{A^2 + M_1^2 \tan^2 \alpha}}, ~~~ 
\cos \beta = {A \over \sqrt{A^2 + M_1^2 \tan^2 \alpha}}, ~~~ 
\end{equation}
and
\begin{equation}
\tan 2 \theta = |B| \cos \alpha {\sqrt{A^2 + M_1^2 \tan^2 \alpha} \over AM_1}.
\end{equation}
If $\alpha=0$, there is no CP violation.  On the other hand, if
$ A^2 << M_1^2 \tan^2 \alpha, $
then
\begin{equation}
e^{i \beta} = {A \over M_1 \tan \alpha} - i, ~~~ 
e^{2i \beta} = -1 - {2iA \over M_1 \tan \alpha}.
\end{equation}
The mass eigenvalues become 
$(M_1 - A \cos 2 \theta - |B|\sin 2 \theta \sin \alpha) e^{2i \delta}$, 
and $-(M_1 + A\cos 2 \theta + |B|\sin 2 \theta \sin \alpha) e^{-2i \delta}$, 
where $\delta = (c^2 A/M_1 \tan \alpha) + (sc|B|\cos \alpha/M_1)$.  Absorbing 
their phases into the definitions of the mass eigenstates $\psi_1, \psi'_1$, 
we then have
\begin{eqnarray}
\psi_1 &=& {e^{i \delta} \over \sqrt 2} \left[ (c e^{-i \beta} - s)N_1 - 
(c e^{-i \beta} + s) N^c_1 \right], \\
\psi'_1 &=& {ie^{-i \delta} \over \sqrt 2} \left[ (c e^{i \beta} + s)N_1 + 
(c e^{i \beta} - s) N^c_1 \right].
\label{psi}
\end{eqnarray}
As a result, the self-energy contributions to the decay of $\psi_1$ through 
$N_1$ and $N^c_1$ with $\psi'_1$ as an intermediate state generate a lepton 
asymmetry
\begin{eqnarray}
\epsilon_1&=& \frac{1}{64\pi}{Im [e^{-4i \delta} (c e^{i \beta} - s)^2 
(c e^{i \beta} + s)^2] \over \Delta M_1/M_1} \left( 
{ 4(\sum_\alpha |h_{\alpha 1}|^2)^2 - (\sum_\alpha |h'_{\alpha 1}|^2)^2 
\over 2\sum_\alpha |h_{\alpha 1}|^2 + \sum_\alpha |h'_{\alpha 1}|^2} 
\right) \nonumber \\ &=& -\frac{1}{64\pi}
 {|B|^2 \sin \alpha \cos \alpha \over A^2 + |B|^2 \sin^2 \alpha} 
 \left( 
{ 4(\sum_\alpha |h_{\alpha 1}|^2)^2 - (\sum_\alpha |h'_{\alpha 1}|^2)^2 
\over 2 \sum_\alpha |h_{\alpha 1}|^2 + \sum_\alpha |h'_{\alpha 1}|^2} \right).
\end{eqnarray}
This is a novel mechanism because there is no CP violation in the Yukawa 
couplings.  Instead it comes from the Majorana phase of the 
$(N_1,N^c_1)$ mass matrix.  Unfortunately, the out-of-equilibrium condition 
for $\psi_1$ decay requires both $|h_1|$ and $|h'_1|$ to be of order 
$10^{-7}$; hence the above contribution to the lepton asymmetry is negligible 
and cannot explain the observed baryon asymmetry of the Universe.

We now consider the contributions of $(N_2,N^c_2)$ to the lepton asymmetry 
in the decays of $\psi_1$ and $\psi'_1$.  We note first that 
if $m'_{11} = m_{11} = 0$
($m'_{22} = m_{22} = 0$),  so 
that $N_1,N^c_1$ ($N_2,N^c_2$) combine to form an exactly Dirac 
fermion, then there is no 
contribution to the asymmetry because lepton number is not broken 
in that system.  
For simplicity let $m'_{11}=m_{11}$ so that the two mass eigenvalues are 
$-M + m_{11}$ and $M + m_{11}$ corresponding to the two mass eigenstates 
$i(N_1-N^c_1)/\sqrt 2$ and $(N_1 + N^c_1)/\sqrt 2$ respectively.  The 
$(N_2,N^c_2)$ system is described by Eqs.~(\ref{mass}) to (\ref{psi}) 
replacing $``1$'' by $``2$''.
The total lepton asymmetry in $\psi_1$ and $\psi'_1$ decays receives 
contributions from the $h$ interactions by themselves, the $h'$ interactions 
by themselves, as well as their interference.  The general expression is 
long and, for the sake of simplicity,  we present only the self-energy 
$h'$ contribution.  [Note that neutrino mass does not depend on $h'$.] 
\begin{eqnarray}
\epsilon_1^{h'} &=&{\Delta M_{1} \Delta M_{2}\over 16 \pi (M_2^2-M_1^2)^3} 
 \\  &\times&
{Im 
\left [
\sum_\alpha (h'_{\alpha 1} h^{'*}_{\alpha 2})^2 
\left( 
\exp (-2i\theta_2) (M_1^4+6 M_1^2 M_2^2 +M_2^4) -
(\sin 2\theta_2/\tan\alpha_2)
(M_1^4-M_2^4)
\right)
\right] \over 
 2 \sum_\alpha |h_{\alpha 1}|^2 + 
\sum_\alpha |h'_{\alpha 1}|^2}. \nonumber
\label{eps}
\end{eqnarray}
Here $\Delta M_1=2m_{11}$, $\Delta M_2=2 A_2/\cos 2 \theta_2,$
and we have expanded assuming $\Delta M_1 << M_1$, $\Delta M_2 << M_2$.
Note that CP violation is again not necessary in the Yukawa couplings. 
Whereas this asymmetry is suppressed by $\Delta M_1\Delta M_2$, it is 
unsuppressed by $h'_{\alpha 2}$ and enhanced by $(M_2^2-M_1^2)^3.$ 
The value $\epsilon_1\sim 10 ^{-6}$ needed to explain 
the observed baryon asymmetry of the Universe may  be obtained consistently
with the neutrino mass parameters and $\Delta a_\mu$ in three ways.

1) If $M_1 \simeq M_2$, there is a resonance factor which comes at the cubic 
power, not in the 
first power as in the canonical leptogenesis. This enhancement factor 
can easily compensate the 
$\Delta M_{1,2}/M_{1,2}$ and Yukawa coupling suppressions. It works 
essentially as 
ordinary resonant leptogenesis with larger asymmetries due to 
more freedom from the $h'$ couplings.


2) If $M_2 >> M_1,$  successful  leptogenesis can be achieved in 
a nonresonant way
provided 
$m_{11}$ or $m'_{11}$ is not much smaller than ${M_1}$. 
Phenomenologically $m_{11}$ and $m'_{11}$ are not constrained by the 
neutrino mass measurements and can be large.
A set of parameters satisfying all constraints 
is for example: $M_1= 2$~TeV, 
$M_2= 5$~TeV, $\Delta M_1/M_1\sim {\cal O}(1)$, 
$\Delta M_2/M_2\sim {\cal O}(10^{-4})$, $h_{2} \sim {\cal O}(10^{-2})$, 
$h'_{2} \sim {\cal O}(10^{-1})$, $h_{1}\sim h'_{1}<10^{-7}$, 
$\lambda_5 \sim {\cal O}(10^{-2})$. [Nonresonant leptogenesis 
at TeV scale from $h'$ type coupling with just 3 right-handed neutrinos (i.e. 
no $\alpha$ phases) has been considered in Ref.~\cite{fhm}.]


3) Finally we may replace $(N_1,N^c_1)$ in Table 1 with a single neutral 
fermion $S$ with zero lepton number, leaving the heavier generations 
unchanged.  The soft breaking of lepton number allows $S$ to mix slightly 
with $(N_{2,3},N^c_{2,3})$ and the decay of $S$ with very much suppressed 
Yukawa couplings is another viable leptogenesis scenario \cite{mss06}.


%
%

Our model can be directly verified at LHC experiments by discovering $N_i$, 
$N^c_i$, $(\eta^+,\eta^0)$ and $\chi^+$ which all have ${\cal O}(10^{-1})$ 
couplings except $N_1$ and $N^c_1$.

In conclusion, we have shown that dark matter, $\Delta a_\mu$, neutrino 
mass, and leptogenesis may all be attributed to particles which are 
{\it odd} under an exact $Z_2$. We have presented an explicit model 
in which $\Delta a_\mu$ is strongly enhanced, thus explaining the experimental 
result without unnaturally large Yukawa couplings. The 
observed neutrino masses are radiatively induced at the TeV scale,  
with extra suppression factors compared to the canonical seesaw: 
a loop factor, {\it small soft} lepton number breaking terms, and an 
extra power of the heavy scale. There are no constraints on neutrino mass 
or $\Delta a_\mu$ coming from $\mu\to e\gamma$. Novel and successful 
leptogenesis at the TeV scale occurs from $(N_1,N^c_1)$ decays due to 
the interference with $(N_{2,3},N^c_{2,3})$. It does not require CP 
violation from  Yukawa couplings and can be induced by new phases coming 
from the $(N_i,N^c_i)$ mass matrices.  It predicts however a (nearly) massless 
neutrino.  Dark matter in the form of $\eta^0$ as well as all the other 
new particles (except $N_1, N^c_1$ because of their small couplings) in 
Table 1 can be directly produced and discovered at the 
LHC or at other future accelerators.  Although there are certainly other 
possibilities to explain the considered phenomenology with help of 
particles odd under the discrete symmetry associated 
with dark matter, our model can be considered as an explicit and minimal 
example of a more general class of TeV scale models. 

~
This work was supported in part by the U.~S.~Department of Energy under Grant 
No.~DE-FG03-94ER40837 (E.M.), by ESF under grant No. 6140 (K.K., M.R.) and 
by a Ramon 
y Cajal contract of the Ministery of Educaci\'on y Ciencia (T.H.).

\newpage
\bibliographystyle{unsrt}

\end{document}